\documentclass[12pt]{JHEP3}
\usepackage{amsmath,amssymb,epsfig,bm,amsfonts,yfonts}
\usepackage{cite}

\newcommand{\be}{\begin{equation}}
\newcommand{\ee}{\end{equation}}
\newcommand{\beqa}{\begin{eqnarray}}
\newcommand{\eeqa}{\end{eqnarray}}
 \newcommand{\bn}{\begin{enumerate}}
\newcommand{\en}{\end{enumerate}}

\def\bra#1{\left\langle #1\right|}
\def\eeq{\end{equation}}

\def\ket#1{\left| #1\right\rangle}

\def\Tr{\mathop{\rm Tr}}

\relax

\title{Small-{\boldmath $x$} single-particle distributions in jets from the coherent branching formalism}

\author{
Sebastian Sapeta\\ 

M. Smoluchowski Institute of Physics,  Jagellonian University,\\
   Reymonta 4, 30-059 Cracow, Poland\\
E-mail : {\tt sapeta@th.if.uj.edu.pl}
}
\author{Urs Achim Wiedemann\\

Department of Physics, CERN, Theory Division, \\ 
CH-1211 Geneva 23, Switzerland\\
E-mail : {\tt Urs.Wiedemann@cern.ch}
}

\abstract{We calculate single parton distributions inside quark and gluon jets
within the coherent branching formalism, which resums leading and
next-to-leading logarithmic contributions. This formalism is at the
basis of the modified leading logarithmic approximation (MLLA), and it
conserves energy exactly. For a wide preasymptotic range of the
evolution variable $Y= \ln \left[ E\theta / Q_0 \right]$, we find marked
differences in the shape and norm of single parton distributions
calculated in the MLLA or in the coherent branching formalism,
respectively. For asymptotically large values $Y \geq 5 - 10$, the
difference in norm persists, while differences in shape disappear. In
this way, our numerical study delineates the jet energy scale needed for
a reliable application of both approaches. We also study the dependence
of the single parton distributions on the hadronization scale $Q_0$ and on 
$\Lambda_{\rm QCD}$, and we
calculate within the coherent branching formalism the identified quark
and gluon distributions inside quark and gluon jets.
}

\preprint{CERN-PH-TH/2008-197}

\begin{document}
\def\vev#1{\langle#1\rangle}
\def\ov{\over}
\def\le{\left}
\def\ri{\right}
\def\ha{{1\over 2}}
\def\lam{{\lambda}}
\def\Lam{{\Lambda}}
\def\al{{\alpha}}
\def\ket#1{|#1\rangle}
\def\bra#1{\langle#1|}
\def\vev#1{\langle#1\rangle}
\def\det{{\rm det}}
\def\tr{{\rm tr}}
\def\Tr{{\rm Tr}}
\def\NN{{\cal N}}
\def\th{{\theta}}

\def\Om{{\Omega}}
\def \th{{\theta}}

\def \lam {\lambda}
\def \om {\omega}
\def \ra {\rightarrow}
\def \ga {\gamma}
\def\sig{{\sigma}}
\def\ep{{\epsilon}}
\def\apr{{\alpha'}}
\newcommand{\p}{\partial}
\def\LL{{\cal L}}
\def\HH{{\cal H}}
\def\GG{{\cal G}}
\def\TT{{\cal T}}
\def\CC{{\cal C}}
\def\OO{{\cal O}}
\def\PP{{\cal P}}
\def\tir{{\tilde r}}

\newcommand{\bea}{\begin{eqnarray}}
\newcommand{\eea}{\end{eqnarray}}
\newcommand{\nn}{\nonumber\\}

\section{Introduction}
Since the early days of QCD, it has been known that destructive interference between soft gluon emissions within a jet suppresses hadron production at small values of the momentum fraction 
$x = p_h/E$. For inclusive single-parton distributions, this is seen already
in the double logarithmic approximation (DLA), which predicts a hump-backed 
plateau~\cite{BasicsQCD}.
However, if one considers the kinematic regime of sufficiently small
momentum fractions $x$, and sufficiently large jet energies $E$, where 
$\ln \left[1/x\right] \sim \ln \left[E/Q_0\right] \sim {\cal O}\left(1/\sqrt{\alpha_s}\right)$, then
corrections to the asymptotic DLA result are of relative order $\sqrt{\alpha_s}$ for
the peak position of the hump-backed plateau~\cite{BasicsQCD, Mueller:1982cq,Mueller:1983cq}. 
These corrections remain sizable up to the highest experimentally accessible jet energies. 
The coherent parton branching formalism~\cite{Bassetto:1984ik, Dokshitzer:1987nm}  leads to 
evolution equations for the inclusive single- and  multi-parton intra-jet distributions, 
which contain the complete set of next-to-leading ${\cal O}(\sqrt{\alpha_s})$ 
corrections, as well as a subset of higher order corrections. The modified leading logarithmic 
approach (MLLA) \cite{Dokshitzer:1987nm,Fong:1990nt} to inclusive parton distributions, which is 
at the basis of many phenomenological comparisons, can be obtained from  
these evolution equations after further approximations. 

One may ask whether the evolution equations of the coherent branching formalism 
contain more physics than MLLA. 
This idea is not supported by parametric considerations, since both MLLA and the 
original evolution equations are complete up to the same order in $\sqrt{\alpha_s}$.
However, the idea is supported by kinematic considerations, since the original evolution
equations conserve energy exactly, while MLLA does not. This has motivated 
in recent years several works which aim at going beyond MLLA. 
In particular, in an approach referred to as NMLLA \cite{Dremin:1994bj, Arleo:2007wn, PerezRamos:2007cr},
one keeps all terms of the original evolution equations, which are one order
${\cal O}(\sqrt{\alpha_s})$ higher than the MLLA accuracy. 
In the same manner also ${\cal O}(\alpha_s)$  corrections to MLLA have been calculated \cite{Dremin:1999ji, Capella:1999ms}.
On the other hand, one may solve
the original evolution equations numerically without any further approximation. So far, this has
been done for fully integrated partonic jet multiplicities only \cite{Lupia:1997in, Lupia:1997bs}. The main result of the present work is 
to solve these original evolution equations for the inclusive single-parton distributions, and
to compare the numerical results to those of MLLA. 

 In the remainder of this introduction, we recall two issues which come up if one 
compares data to the QCD calculations reviewed above. First, the range of applicability 
of the evolution equations is limited to sufficiently high energies, and a full numerical solution 
will allow us to be quantitative about the scale above which a comparison of the formalism 
with data may be regarded as being reliable. Second, it has been emphasized repeatedly
that the QCD prediction for the hump-backed plateau of the single inclusive distribution remains
unaffected by the non-perturbative hadronization process as long as hadronization is sufficiently 
local so that it does not alter significantly the shapes of partonic momentum distributions. In 
particular, the phenomenologically successful comparison of the (rescaled) partonic MLLA 
prediction with inclusive hadronic distributions has been viewed as support for a local 
parton-hadron duality (LPHD) underlying the hadronization 
mechanism~\cite{Azimov:1984np,Azimov:1985by}. On the other
hand, hadronization models employed e.g. in Monte Carlo event generators, such as
the Lund string fragmentation or cluster hadronization lead to a significant further 
softening of single-inclusive distributions during 
hadronization~\cite{Sjostrand:2006za,Marchesini:1991ch,Gleisberg:2003xi}. Any 
difference between MLLA, NMLLA and the exact numerical solution to the evolution 
equations in the coherent branching formalism may point to the possible relevance of 
further higher-order and/or non-perturbative effects, and may thus add to our picture of 
the hadronization process.

The paper is organized as follows. In Sec.~\ref{sec2},  we introduce the equations of the coherent branching formalism~\cite{Bassetto:1984ik, Dokshitzer:1987nm} and we define all quantities of interest. In particular, we discuss in Sec.~\ref{sec:init} the general ansatz for the initial conditions  of these
equations. The MLLA approach is briefly recalled in Sec.~\ref{sec:mlla}. In Sec.~\ref{sec:fulleq} we describe how we solve the full evolution equations of the coherent branching formalism. 
The numerical solutions are discussed in Sec.~\ref{sec:numres}. 
In Sec.~\ref{sec3a}, we present the single-particle spectra in the phenomenologically accessible as well as in the asymptotic range of jet energies. In Sec.~\ref{sec3b} we show the results for total multiplicity and total energy of partons. In Sec.~\ref{sec3c} we study the dependence of these results on the
hadronization scale. 
In Sec.~\ref{sec:maching}, we explore finally to what extent characteristic differences between
the coherent branching formalism and MLLA remain, if one allows for the freedom of adapting
the norm and hadronization scale independently in both approaches. Our method of solving the equation of the coherent branching formalism enables us also to determine separately the distributions of quarks and gluons in a quark or a gluon jet. The corresponding results are discussed in Sec.~\ref{sec:idpartons}. 
Our conclusions are summarized in Sec.~\ref{sec:conclusion}.

\section{Evolution equations for single parton distributions at small {\boldmath $x$}}
\label{sec2}

We want to calculate the single distributions $D_q(x,Y)$ and $D_g(x,Y)$ of partons in 
a quark or gluon jet, respectively. Here, $x = p/E$ is the 
jet momentum fraction of partons, and the variable
\begin{equation}
\label{eq:ydef}
 Y = \ln\frac{E \theta}{Q_0}
 \end{equation}
 is written in terms of the sufficiently small jet opening angle  $\theta$, and the hadronic scale 
 $Q_0$. We denote the logarithm of the momentum fraction by
\begin{equation}
	l = \ln \frac{1}{x}\, .
\end{equation}
In the coherent parton branching formalism, the evolution equations for these distributions are given by~\cite{Bassetto:1984ik, Dokshitzer:1987nm}
\bea
\label{eq:evoleqs}
\partial_Y D_q(x,Y) & = &
\frac{1}{2}\int_{z_-}^{z_+} dz \, \frac{\alpha_s(k^2_{\perp})}{\pi}\, 
\nonumber \\
& &
\hspace{1.0cm}
\times \Bigg\{
P_{qq}(z)\,
\bigg[D_q\Big(\frac{x}{z},Y+\ln z\Big) + 
       D_g\Big(\frac{x}{1-z},Y+\ln (1-z)\Big) - 
       D_q(x,Y)\bigg] \nonumber \\
& &
\hspace{1.2cm}
+\, P_{gq}(z)\,
\bigg[D_g\Big(\frac{x}{z},Y+\ln z\Big) + 
       D_q\Big(\frac{x}{1-z},Y+\ln (1-z)\Big) - 
       D_q(x,Y)\bigg] 
\Bigg\}, \nonumber \\
\partial_Y D_g(x,Y) & = &
\frac{1}{2}\int_{z_-}^{z_+} dz \, \frac{\alpha_s(k^2_{\perp})}{\pi}\, 
\nonumber \\
& &
\hspace{1.0cm}
\times \Bigg\{
P_{gg}(z)\,
\bigg[D_g\Big(\frac{x}{z},Y+\ln z\Big) + 
       D_g\Big(\frac{x}{1-z},Y+\ln (1-z)\Big) - 
       D_g(x,Y)\bigg] \nonumber \\
& &
\hspace{1.0cm}
+\, 2 n_f \, P_{qg}(z)\,
\bigg[D_q\Big(\frac{x}{z},Y+\ln z\Big) + 
       D_q\Big(\frac{x}{1-z},Y+\ln (1-z)\Big) - 
       D_g(x,Y)\bigg] 
\Bigg\}, \nonumber  \\ 
& & \nonumber  \\ 
\eea
with the unregularized splitting functions 
\bea
P_{qq} (z) & = & C_F \frac{1+z^2}{1-z}, \nonumber \\
P_{gg} (z) & = & 2C_A 
\left(\frac{1-z}{z}+\frac{z}{1-z}+z(1-z)\right), \nonumber \\
P_{qg} (z) & = & T_R \left(z^2 + (1-z)^2\right), 
\quad T_R = \frac{1}{2},
\nonumber \\
P_{gq} (z) & = & C_F \frac{1+(1-z)^2}{z}.
\eea

To calculate jet multiplicity distributions with next-to-leading logarithmic accuracy, one can use a DGLAP chain of $1 \to 2$ parton branchings which follows an exact angular ordering prescription (in contrast to the strong angular ordering prescription in DLA) and in which the coupling constant depends on $k_\perp^2$ at each vertex.  In the set of evolution equations (\ref{eq:evoleqs}) the angular ordering is implemented by the choice of the arguments $(Y+\ln z)$ and $(Y+ \ln (1-z))$ of the parton distributions and also the $k_\perp$ prescription is applied for the coupling~\cite{Bassetto:1984ik}.

\begin{figure}[t]
\begin{center}
\includegraphics[height=16.0cm,angle=-90]{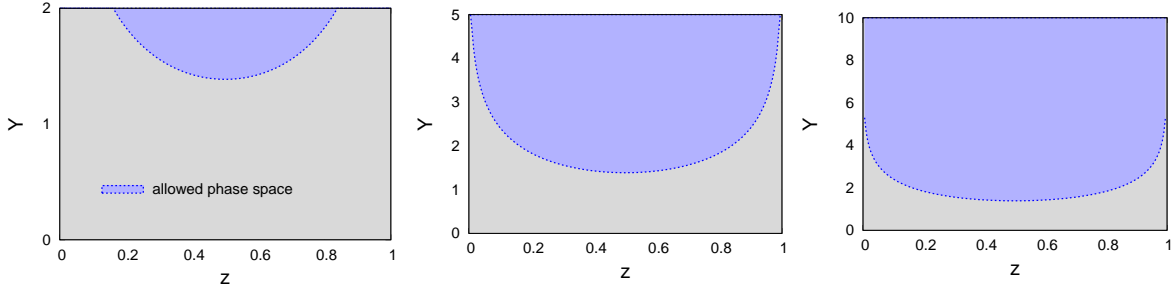}
\end{center}
\caption{The range (\protect\ref{2.6}) of $z$-values, which are kinematically
allowed in the evolution of the single-parton distributions $D_q(x,Y)$, 
$D_g(x,Y)$ up to $Y=2$ (left), $Y=5$ (middle), $Y=10$ (right).}
\label{fig:cutoff}
\end{figure}

The limits $z_-$, $z_+$ on the $z$-integral in (\ref{eq:evoleqs}) are set by the 
requirement that the transverse momentum $k_{\perp}$ is sufficiently large for 
perturbative evolution to be valid. That means, $k_{\perp}$ is larger than a
hadronic scale $Q_0$
\be
\label{2.5}
k_{\perp} \approx z(1-z)E\theta \geq Q_0
\quad \Rightarrow \quad Y + \ln z + \ln (1-z) \geq 0.
\ee
For $Y < \ln 4$, this inequality is not satisfied for any real $z$, so evolution occurs only 
for $Y> \ln 4$, where the inequality is fulfilled for
\be
\label{2.6}
z_- \equiv \frac{1}{2}\Big(1-\sqrt{1-4 e^{-Y}}\Big) < z <
z_+ \equiv \frac{1}{2}\Big(1+\sqrt{1-4 e^{-Y}}\Big). 
\ee
This kinematical regime is depicted in Fig.~\ref{fig:cutoff}.
%
%
One can check that if (\ref{2.6}) is satisfied, then the arguments 
$\left(Y + \ln z\right)$ and $\left( Y + \ln (1-z)\, \right)$ of the functions $D_q$ and $D_g$ 
in  (\ref{eq:evoleqs}) cannot be negative. 
The coupling constant can be written as
\be
\alpha_s(k^2_{\perp})=\frac{\pi}{2 N_c \, \beta}\, \frac{1}{Y+\ln z+\ln (1-z) + \lambda},
\quad
\beta = \frac{1}{4N_c}\left(\frac{11}{3}N_c - \frac{4}{3}n_f T_R\right),
\quad
\lambda = \ln\frac{Q_0}{\Lambda_{\rm QCD}}.
\label{2.7}
\ee
%
%
\subsection{Initial conditions and ansatz for solution}
\label{sec:init}
A parton shower in the real world is pictured often as a quark or a gluon produced 
initially with high energy and virtuality and evolving perturbatively from large $Y$ to 
a small hadronic scale $Y_0$. Evolution occurs by emitting partons at smaller and 
smaller angles, till a minimal angle $\theta_0$ is reached, at which 
non-perturbative hadronic effects set in. The numerical solution of the evolution 
equations (\ref{eq:evoleqs}) will proceed in the opposite direction, that is, the initial
conditions for the solution of (\ref{eq:evoleqs}) are set at the small scale $Y_0$ which 
corresponds to the final partonic state of the physical process. These initial conditions specify
which partons are measured. For instance, consider the initial condition
\bea
\label{eq:init1}
D_q(x,Y_0)  =  \delta(1-x)\, ,\qquad 
D_g(x,Y_0)  =  0\, .  \qquad \hbox{(Case I)}
\eea
This initial condition for the evolution equation specifies that the partons at the low 
scale $Y_0$ cannot split further and, by virtue of the choice
(\ref{eq:init1}), these partons are quarks. Starting the evolution from the initial condition 
(\ref{eq:init1}), the functions $D_q(x,Y)$ and $D_g(x,Y)$ denote the single-{\it quark} distributions  in a
quark and a gluon jet, respectively. Alternatively, for the initial condition
\bea
\label{eq:init2}
D_q(x,Y_0)  =  0\, ,\qquad 
D_g(x,Y_0)  =  \delta(1-x)\, ,  \qquad \hbox{(Case II)}
\eea
the functions $D_q(x,Y)$ and $D_g(x,Y)$ denote the single-{\it gluon} distributions in a
quark and a gluon jet, respectively. Previous studies \cite{Dokshitzer:1987nm,Dremin:1994bj, Fong:1990nt, Arleo:2007wn, PerezRamos:2007cr,Dremin:1999ji, Capella:1999ms,Lupia:1997in,Lupia:1997bs}
have focused on the
distribution of {\it all} partons in a quark or a gluon jet. This is obtained by evolving
 $D_q(x,Y)$ and $D_g(x,Y)$ from the initial condition
\bea
\label{eq:init3}
D_q(x,Y_0)  =   \delta(1-x)\, ,\qquad 
D_g(x,Y_0)  =  \delta(1-x)\, . \qquad \hbox{(Case III)}
\eea
In Sec.~\ref{sec:numres} of this paper we present the results for the case of the initial condition (III) whereas in Sec.~\ref{sec:idpartons} we consider cases (I) and (II).

The most general ansatz for the solutions of the evolution equations (\ref{eq:evoleqs}) reads
\bea
D_q(l,Y) & = & S_q(Y)\,\delta(l) + Q(l,Y),  
\label{2.11} \\
D_g(l,Y) & = & S_g(Y)\,\delta(l) + G(l,Y). 
\label{2.12}
\eea
Here, the discrete parts are expressed in terms of the Sudakov form factors $S_q(Y)$ and 
$S_g(Y)$, which denote the probability that a parton does not split in the evolution between 
$Y_0$ and $Y$. These discrete terms arise as a natural consequence of the initial
conditions (\ref{eq:init1}), (\ref{eq:init2}) and (\ref{eq:init3}).
The continuous parts $Q(l,Y)$ and $G(l,Y)$ vanish at the initial scale $Y=Y_0$ and arise
from the further evolution with (\ref{eq:evoleqs}).

\subsection{The MLLA evolution equations}
\label{sec:mlla}

In the following sections, we shall compare the numerical solution of (\ref{eq:evoleqs}) to 
the results obtained in the MLLA approach. Here, we recall the main approximations involved 
in deriving MLLA from the Eqs.~(\ref{eq:evoleqs}). For more details about MLLA, we refer to \cite{Dokshitzer:1987nm, BasicsQCD}.

In the limit of large $Y$ and small $x$  one can perform an expansion of Eqs.~\eqref{eq:evoleqs} in powers of $\sqrt{\alpha_s}$. Formally, this is done by treating logarithms $Y$ and
$\ln 1/x$ as being of order $1/\sqrt{\alpha_s}$. 
Keeping only terms of order  ${\cal O}(\sqrt{\alpha_s})$ on the right hand side of 
Eqs.~\eqref{eq:evoleqs} (i.e. keeping terms of order ${\cal O}(1)$ in the distributions $D_q$ and 
$D_g$), these equations reduce to the DLA evolution equations with running coupling. 
The subleading corrections to DLA are obtained by keeping terms up to ${\cal O}(\alpha_s)$ on the 
right hand side of (\ref{eq:evoleqs}) (thus keeping terms of order ${\cal O}(\sqrt{\alpha_s})$ in  
$D_q$ and $D_g$). This leads to~\cite{Ramos:2006dx}
\bea
\label{2.13}
\partial_Y D_q(x,Y) & = &
\frac{C_F}{N_c}
\left\{ 
\int^1_0 \frac{dz}{z} 
\, \gamma_0^2(Y+\ln z)\, D_g\Big(\frac{x}{z},Y+\ln z\Big) -
a_{qg} \, \gamma_0^2(Y)\, D_g\left(x,Y\right)
\right\}, \hspace{1.0cm} \\
\label{2.14}
\partial_Y D_g(x,Y) & = &
\int^1_0 \frac{dz}{z} 
\, \gamma_0^2(Y+\ln z)\, D_g\Big(\frac{x}{z},Y+\ln z\Big) -
  a_{gq} \, \gamma_0^2(Y)\, D_q\left(x,Y\right)-
  a_{gg} \, \gamma_0^2(Y)\, D_g\left(x,Y\right),
\nonumber \\
\eea
where 
$\gamma^2_0(Y)= 1/(\beta(Y + \lambda))$
and
\be
a_{qg} = \frac{3}{4},
\qquad
a_{gq} = -
\frac{2}{3} \frac{n_f T_R}{N_c},
\qquad
a_{gg} = \frac{1}{4 N_c}
\left(\frac{11}{3} N_c+ \frac{4}{3} n_f T_R \right).
\ee
In deriving Eqs.~\eqref{2.13} and \eqref{2.14}, the semi-hard splittings 
$z\sim 1$ from the evolution equation (\ref{eq:evoleqs}) have been taken into account only
partially. For this reason, energy is not conserved exactly in  
the above equations.

Formally, to arrive at the MLLA equation~\cite{Dokshitzer:1987nm,BasicsQCD}, one requires the DLA relation 
\begin{equation}
   D_q(l,Y)  = \frac{C_F}{N_c} D_g(l,Y)\, .
  \label{2.**}
\end{equation}
By inserting this expression into (\ref{2.14}), one obtains a closed equation for the parton
distribution in a gluon jet
\be
\label{eq:mlla}
\partial_Y D_g(x,Y)  = 
\int^1_0 \frac{dz}{z} 
\, \gamma_0^2(Y+\ln z)\, D_g\Big(\frac{x}{z},Y+\ln z\Big) -
 a_1 \, \gamma_0^2(Y)\,D_g\left(x,Y\right),
\ee
where
\be
a_1 = a_{gg}+\frac{C_F}{N_c} a_{gq} =\frac{1}{4 N_c}
\left[
\frac{11}{3} N_c+ \frac{4}{3} n_f T_R \left( 1- \frac{2 C_F}{N_c}\right)
\right].
\ee
Eq.~(\ref{eq:mlla}) is commonly referred to as MLLA evolution equation. 
From this, the parton distribution in a quark jet is usually determined with the help
of (\ref{2.**}). 
Compared to the DLA approximation, the MLLA equation shows mainly two improvements: 
the coupling constant is running and
the negative term $\propto -\gamma_0^2(Y) G\left(x,Y\right)$ on the right hand side of
Eq.~\eqref{eq:mlla} accounts for recoil effects which
lead to a softening of the
spectra as compared to DLA.

We note that the MLLA evolution equation \eqref{eq:mlla} propagates only the initial conditions  (II) or (III), which give the same result for $D_g(x,Y)$.  This is because within the MLLA approach one cannot differentiate between quarks and gluons in a quark or a gluon jet. 
In the following, all numerical results for the MLLA evolution are obtained 
with the initial conditions $D_g(x,Y)=\delta(1-x)$.
Using the ansatz (\ref{2.12}) one obtains from Eq.~\eqref{eq:mlla} the MLLA Sudakov form factor
\begin{equation}
	S_{\rm MLLA}(Y) = \left( \frac{Y+\lambda}{\lambda}\right)^{-a_1/\beta}\, 
	\label{2.18}
\end{equation}
and the equation for the continuous part $G(l,Y)$.

\subsection{Evolution equations in the coherent branching formalism}
\label{sec:fulleq}

\begin{figure}[t]
\begin{center}
\includegraphics[height=17.0cm, angle=-90]{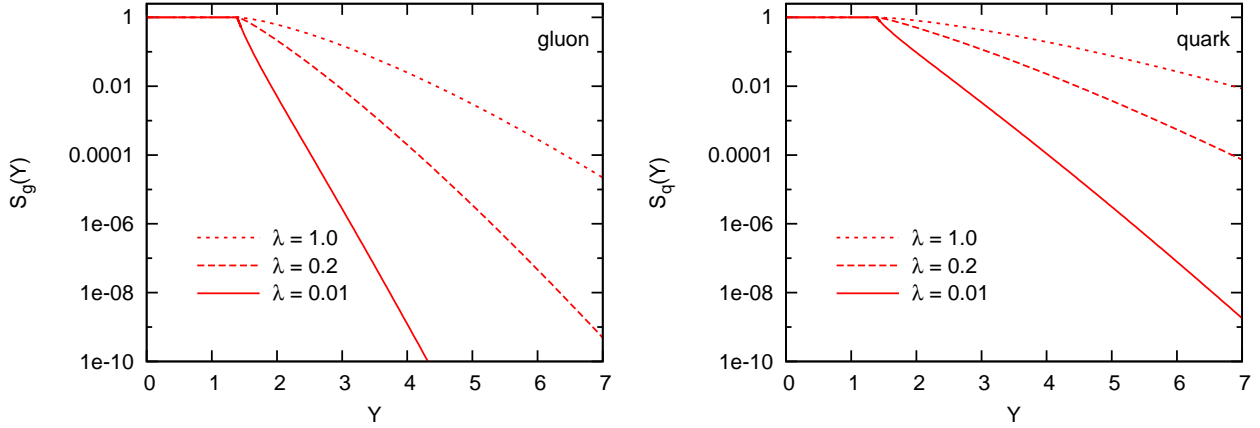}
\caption{The Sudakov form factors (\protect\ref{eq:sudeq1}) and 
(\protect\ref{eq:sudeq2}) for three different values
of $\lambda$.}
\label{fig2}
\end{center}
\end{figure}

In this subsection, we return to the full evolution equations
(\ref{eq:evoleqs}) of the coherent branching formalism. Inserting the ansatz (\ref{2.11}), 
(\ref{2.12}) into (\ref{eq:evoleqs}), we find for the Sudakov form factors 
\bea
\partial_Y S_q(Y) & = & 
-\frac{1}{2}\int^{z_+}_{z_-} dz \, \frac{\alpha_s(k^2_{\perp})}{\pi}\, 
\left[P_{qq}(z)+P_{gq}(z)\right] S_q(Y),
\label{2.19} \\
\partial_Y S_g(Y) & = & 
-\frac{1}{2}\int^{z_+}_{z_-} dz \, \frac{\alpha_s(k^2_{\perp})}{\pi}\, 
\left[P_{gg}(z)+2 n_f P_{qg}(z)\right]  S_g(Y).
\label{2.20}
\eea
The above equations are not coupled and can be easily solved. It follows from the integration boundaries $z_-$, $z_+$ given in (\ref{2.6}), that 
\be
S_q(Y)  =  S_g(Y)  = 1 \qquad {\rm for}\ Y< \ln 4\, . 
\label{2.21}
\ee
After exploiting the $z\leftrightarrow 1-z$ symmetry
of the transverse momentum (\ref{2.5}) and the integration boundaries, and with the help of the property $P_{qq}(z) = P_{gq}(1-z)$, we find for the initial condition~\eqref{eq:init3} and for $Y > \ln 4$
\bea
\label{eq:sudeq1}
S_q(Y) & = & 
\exp\left\{
-\frac{C_F}{2N_c \beta}\int_{\ln 4}^{Y}dY'\int^{z_+}_{z_-} 
\frac{dz}{Y' + \ln z(1-z) + \lambda} \, 
\frac{1+z^2}{1-z}
\right\},
\\
\label{eq:sudeq2}
S_g(Y) & = & 
\exp\Bigg\{
-\frac{1}{4N_c \beta}
\int_{\ln 4}^{Y}dY'\int^{z_+}_{z_-} dz \, 
\frac{dz}{Y' + \ln z(1-z) + \lambda} \, 
\nonumber \\
&&
\qquad
\times
\left[
2C_A 
\left(\frac{1-z}{z}+\frac{z}{1-z}+z(1-z)\right)
+2 n_f T_R \left(z^2 + (1-z)^2\right)
\right] 
\Bigg\}.
\eea
We have performed the double integrals in Eqs.~\eqref{eq:sudeq1} and~\eqref{eq:sudeq2} 
numerically.  Fig.~\ref{fig2} shows the result for three values of $\lambda$.  
For $Y > \ln 4$, these factors decrease almost exponentially 
in sharp contrast to the MLLA behavior \eqref{2.18}. One can show that for large $Y$ the slope changes with $Y$ according to
\be
\frac{d\ln S_{q,g}}{dY}  \simeq 
- c_{q,g}\, \ln \left[\frac{Y+\lambda}{\lambda}\right],
\label{logY}
\ee
where
$c_q = C_F/(N_c \beta)$ and
$c_g = 1/\beta$.

The solutions $S_q(Y)$, $S_g(Y)$ enter the evolution 
equations for $Q(l,Y)$ and $G(l,Y)$. After changing the integration variable $z$ to \mbox{$\bar l = \ln(z/x)$}, the equations for the continuous parts of the spectra can be written as
\bea
\label{eq:conteqs}
\partial_Y Q(l,Y) & = &
\int^{\bar l_{\rm max}}_{\bar l_{\rm min}} d\bar l \, \Gamma_s(l-\bar l, Y)\, 
 \bigg\{
P_{qq}(l-\bar l)\,Q(\bar l,Y-l+\bar l) +
P_{gq}(l-\bar l)\,G(\bar l,Y-l+\bar l) 
\bigg\} \nonumber \\
& &
\hspace{5pt} + \
\Gamma_s(l, Y) \bigg\{
P_{qq}(l)\,S_q(Y-l) +
P_{gq}(l)\,S_g(Y-l) 
\bigg\}  
+ \partial_Y\! \ln S_q(Y) \, Q(l,Y),
\nonumber \\
\partial_Y G(l,Y) & = &
\int^{\bar l_{\rm max}}_{\bar l_{\rm min}} d\bar l \, \Gamma_s(l-\bar l, Y)\, 
 \bigg\{
P_{gg}(l-\bar l)\,G(\bar l,Y-l+\bar l) +
2 n_f P_{qg}(l-\bar l)\,Q(\bar l,Y-l+\bar l)
\bigg\} \nonumber \\
& &
\hspace{5pt} + \
\Gamma_s(l, Y) \bigg\{
P_{gg}(l)\,S_g(Y-l) +
2 n_f P_{qg}(l)\,S_q(Y-l) 
\bigg\}  
+ \partial_Y\!\ln S_g(Y)\, G(l,Y),
\nonumber \\
& & \nonumber  \\ 
\eea
where we have introduced the shorthand
\be
\Gamma_s(l,Y) = 
\frac{1}{2 N_c \beta}\, \frac{e^{-l}}{Y-l +\ln (1-e^{-l})+\lambda}.
\ee
The limits of integration in Eq.~\eqref{eq:conteqs} are
\be
\bar l_{\rm min}=l-\ln\frac{2}{1-\sqrt{1-4 e^{-Y}}}\, ,
\qquad
\bar l_{\rm max} = l-\ln\frac{2}{1+\sqrt{1-4 e^{-Y}}}\, .
\ee
We recall that the functions $G(l,Y)$ and $Q(l,Y)$ vanish when the first argument is negative.
We also note that the form (\ref{eq:conteqs}) for the evolution equations in the 
coherent branching formalism is valid for all initial conditions and the logarithmic derivatives $\partial_Y\! \ln S_{q,g}(l,Y)$ on the right hand side of these
evolution equations are formal notational shortcuts, denoting the expressions
(\ref{eq:sudeq1}), (\ref{eq:sudeq2}) irrespective of the initial conditions.  
These logarithmic derivatives decrease only logarithmically with $Y$,
as seen from Eq.~(\ref{logY}). Hence, the rapid decrease of the Sudakov form factors itself 
does not render the corresponding terms in (\ref{eq:conteqs}) negligible at large $Y$.

\section{Numerical results}
\label{sec:numres}

We have solved numerically the evolution equations for the continuous parts of 
single particle distributions in quark and gluon jets in the coherent parton branching
formalism~\eqref{eq:conteqs} and in the MLLA approach (\ref{eq:mlla}). 
The solutions depend on 
 \begin{equation}
 	\lambda = \ln Q_0/\Lambda_{\rm QCD}\, ,
 \end{equation}
 which specifies in units of $\Lambda_{\rm QCD}$ the value of the hadronization scale 
 $Q_0$ up to which the parton shower is evolved. It is a peculiar feature of the MLLA 
 approach, that its solutions have a well-defined finite limit $\lambda \to 0$, which is known
 as the ``limiting spectrum'' and which is at the basis of many phenomenological 
 comparisons \cite{BasicsQCD,Azimov:1984np,Azimov:1985by,Akrawy:1990ha,Acosta:2002gg}. 
 In general, however, one cannot expect to obtain
 finite results in the limit $\lambda \to 0$, since $\lambda$ regulates the Landau pole of 
 the strong coupling constant (\ref{2.7}). 
In fact, we shall find that the solutions for the evolution 
 equations ~\eqref{eq:evoleqs} can be obtained for a finite value of $\lambda$, only.
 
 In the following, we show numerical results for $\lambda = 0.01$, $\lambda = 0.2$ and 
 $\lambda = 1.0$. The MLLA limiting spectrum is approximately 
$3-4\%$ larger in norm and similar in shape, compared to the MLLA solution for $\lambda = 0.01$. 
Hence, our choice for this smallest value 
 of $\lambda$ is motivated by the interest in exploring the solutions of the coherent branching
 formalism ~\eqref{eq:evoleqs} for a parameter range in which the MLLA limiting spectrum
 is almost reached. Our choice of the largest value $\lambda = 1.0$ is motivated by 
 the fact that parton showers implemented in Monte Carlo event generators typically
 end the perturbative evolution at a hadronization scale $Q_0$ which is significantly
 larger than $\Lambda_{\rm QCD}$ and for which $\lambda \sim 1$ is an order
 of magnitude estimate. We have chosen one value $\lambda = 0.2$ in between, and it is for this
 value that we shall first present our results.
\subsection{The distributions $Q(l,Y)$ and $G(l,Y)$}
\label{sec3a}
%
\begin{figure}[t]
\begin{center}
\includegraphics[height=13.0cm, angle=0]{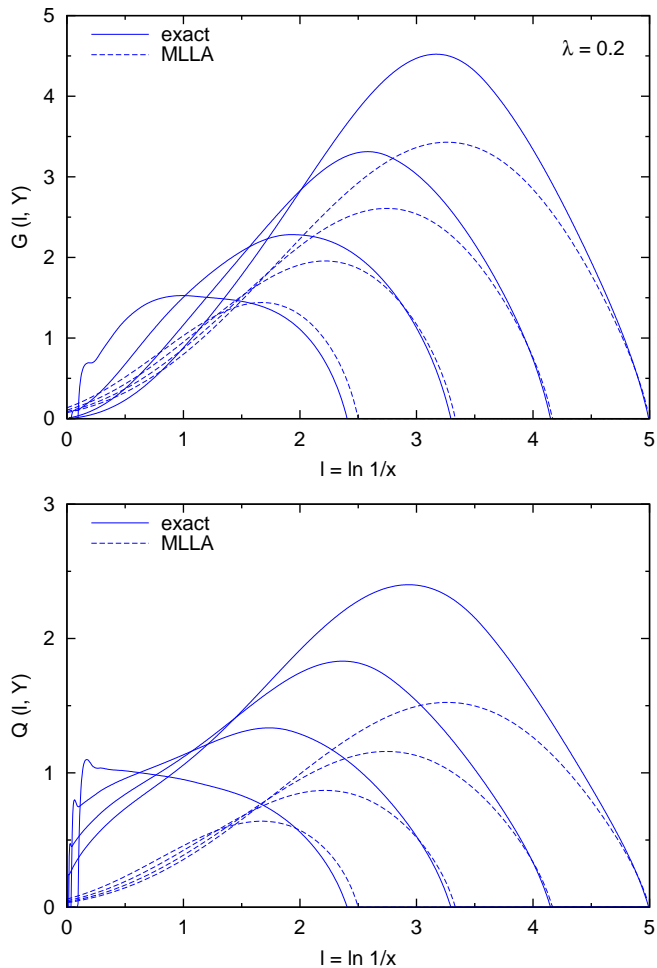}
\caption{The continuous parts of the inclusive distribution of all partons in gluon (upper panel) and  quark jet (lower panel). Results are shown for evolution in the coherent parton branching 
formalism~(\protect\ref{eq:evoleqs}), and in the MLLA approach 
(\protect\ref{eq:mlla}) for $\lambda = 0.2$ and $Y=$ 2.5, 3.3, 4.2 and  5.0.}
\label{fig3}
\end{center}
\end{figure}

\begin{figure}[t]
\begin{center}
\includegraphics[height=13.0cm, angle=0]{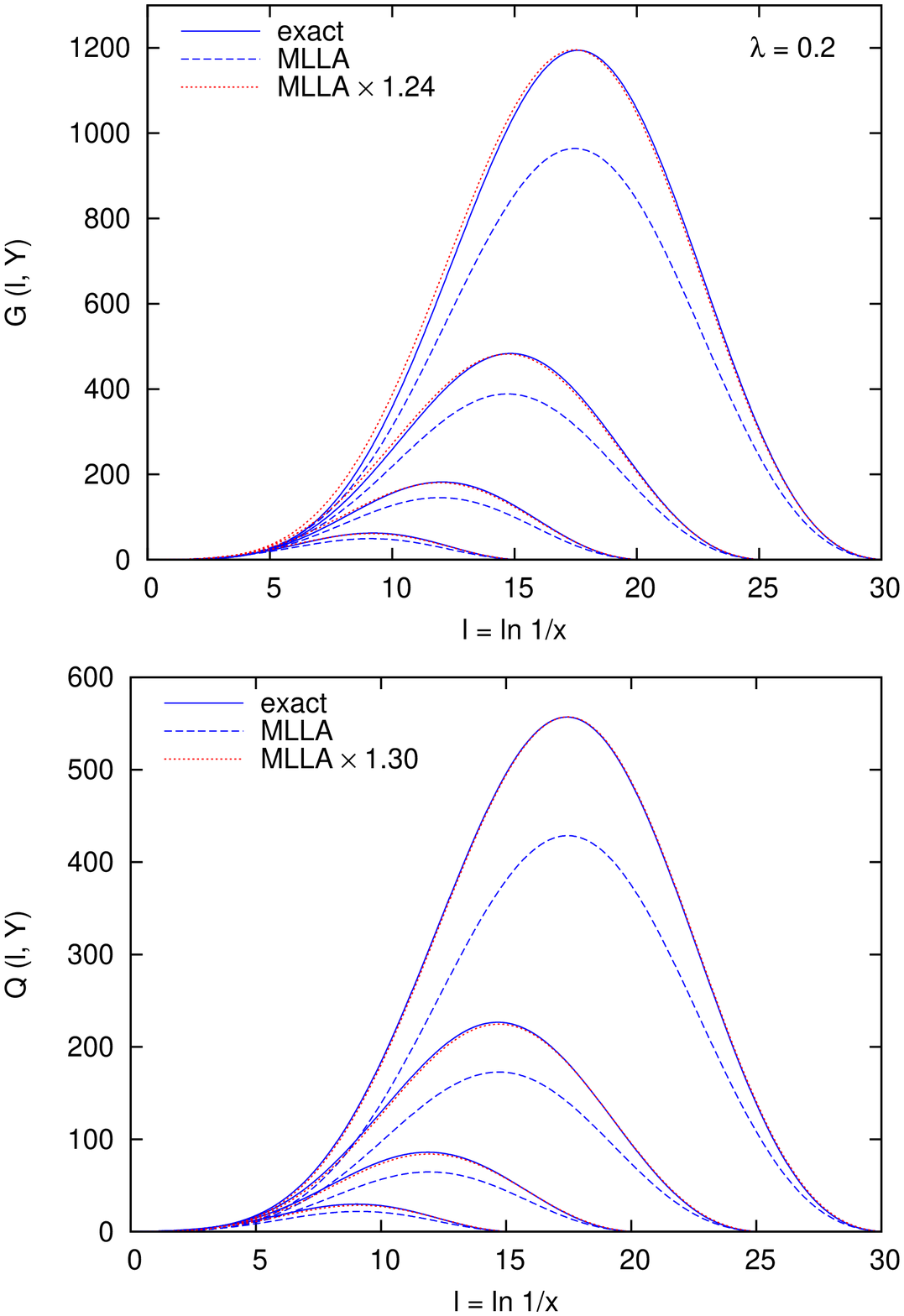}
\caption{Same as Fig.~\protect\ref{fig3}, but for evolution up to $Y=15$, $20$, $25$ and $30$.}
\label{fig4}
\end{center}
\end{figure}

In Figs.~\ref{fig3} and ~\ref{fig4}, we present results for the continuous part of the 
single-parton distributions inside a quark and a gluon jet, evolved from  the initial 
condition (\ref{eq:init3})  at $Y_0=0$ up to different values of $Y$. Fig.~\ref{fig3} shows results for the evolution up to values
of $Y\leq 5$ whereas Fig.~\ref{fig4} focuses on evolution to much larger values $15 \leq Y \leq 30$. 
The relation between the evolution variable $Y$ and the jet energy $E$ depends on the
jet opening angle $\theta$ and the hadronization scale $Q_0$, as seen from Eq.~\eqref{eq:ydef}. 
In the case of large angles, the approximate formula \eqref{eq:ydef} must be replaced by the exact relation $Y= \ln (E \sin\theta/Q_0)$.
For a given jet energy,
the maximal value of $Y$ is obtained with the maximal choice for $\theta$ 
and the minimal choice for $Q_0$. 
For instance, for a maximal angle, $\theta = \pi/2$, and a very low value
$Q_0 = 200$ MeV, the value of $Y = 5$ corresponds to an energy of $E=30$~GeV and
$Y=7$ corresponds to $E=220$~GeV. On the other hand, for a smaller opening $\theta = 0.3$,
a choice of $Y=5$ corresponds to $E=100$~GeV. Also, for $\theta = 0.3$, $Y=5$ corresponds 
to $E=250$~GeV if one chooses a larger hadronization scale $Q_0=500$ MeV. These numerical 
examples illustrate that while the correspondence between jet energy and $Y$ may vary significantly 
depending on the choice of $Q_0$, the range of $Y\leq 5-7$ matches quite well the entire
range of jet energies which have been compared so far to data on inclusive single hadronic 
distributions inside jets. This is one reason for considering first the evolution up to $Y=5$.
We note that similar numerical estimates relate $Y=10$ to a minimal jet energy 
$E>4$~TeV. So, our choice of larger values of $Y$ can be motivated solely by the
theoretical interest of testing the asymptotic behavior of these evolution equations.  
In addition, we recall that the evolution equations \eqref{eq:evoleqs} have been derived in an approximation, in which logarithms
 are large and corrections of order $Y \sim 1/\sqrt{\alpha_s}$ are kept. Starting at $Y_0$,
 some evolution will be necessary to satisfy this condition. As we shall see, this initial
 stage of the evolution leads to some peculiar features in the solution which can persist up to
 values of $Y=5$ or larger.

We consider first the case for $Y<5$. 
As seen in Fig.~\ref{fig3}, the single inclusive distributions obtained for the evolution within the coherent branching formalism 
show for small $Y$ a large yield at low values of $l \ll Y$. This effect is more pronounced for
the distribution in a quark jet, than for the distribution in a gluon jet. It can be understood on
general grounds from the fact that the initial conditions (\ref{eq:init3}) for the evolution
are delta-functions at $l =0$, and that the results for small $Y$ remember these initial 
conditions. Moreover, quarks are less likely to branch than gluons, and in their first branching 
they are likely to keep a large fraction of their energy. This tends to make the distribution in a 
quark jet harder than in a gluon jet, as clearly seen in Fig.~\ref{fig3}. Upon further evolution
to $Y=3.3, 4.2, 5.0$, the yield of hard partons (say partons with $l<1$) in a gluon jet decreases
rapidly to a value comparable with (and for some region of $l$ even smaller than) the results 
obtained in the MLLA approach. On the other hand, the yield of hard partons in a quark jet
remains enhanced significantly over the yield obtained in the MLLA approach up to 
relatively large values of the evolution parameter $Y$. We recall that the parametric
reasoning underlying the evolution equations \eqref{eq:evoleqs} is valid for sufficiently
large $Y$ and $l \sim {\cal O}(Y)$. However, also in the region around the peak of the
single inclusive distributions (which will lie for sufficiently large $Y$ within this kinematic
range of validity), there are significant differences between the MLLA approach and the
coherent branching formalism.

\begin{figure}[t]
\begin{center}
\includegraphics[height=10.0cm, angle=-90]{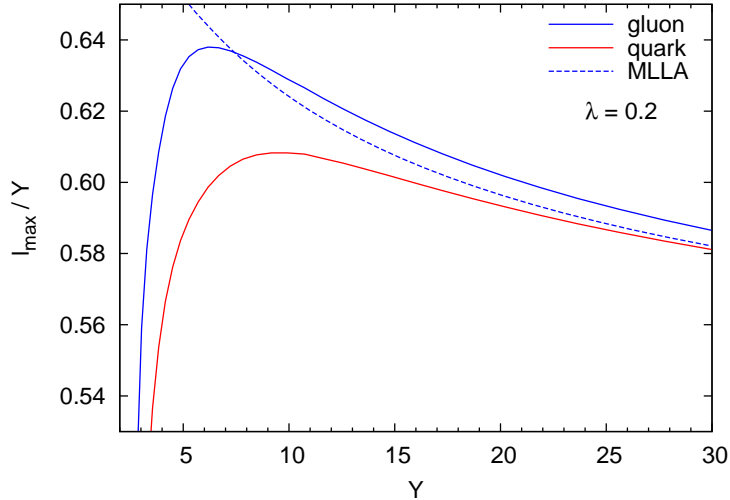}
\caption{The evolution in $Y$ of the peak position of the continuous parts of the 
inclusive parton distributions. Results are shown for all partons in a gluon jet
(upper straight line) or a quark jet (lower straight line), calculated  in the
coherent parton branching formalism~(\protect\ref{eq:evoleqs}). Results are also shown
for the MLLA approach  (\protect\ref{eq:mlla}) (dashed line), for which the peak positions 
of the distributions for quark and gluon jets coincide. All results are calculated for 
$\lambda = 0.2$ and 
the peak position $l_{\rm max}$ is plotted in units of $Y$.}
\label{fig5}
\end{center}
\end{figure}

From Fig.~\ref{fig3}, we have seen several preasymptotic features, which become less 
pronounced upon further evolution in $Y$. Fig.~\ref{fig4} illustrates the degree to which
these preasymptotic features disappear with increasing $Y$. The numerical results
for the coherent branching formalism and the MLLA approach agree very well in shape
and differ in norm by a factor of order unity which is $Y$-independent to good accuracy.
Closer investigation shows that the factors needed to adjust the norm between both 
approaches differ for gluon jets (normalization factor 1.24 for $\lambda = 0.2$) and 
quark jets (normalization factor 1.30 for $\lambda = 0.2$). 
Fig.~\ref{fig4} illustrates the statement that in the asymptotic region of evolution towards
sufficiently large $Y$, the region around the peak of the single-parton distributions lies 
at sufficiently large $l$ to be calculable within the accuracy of the single branching 
formalism. In this asymptotic region, MLLA and coherent branching formalism agree
up to the norm. On the other hand, Fig.~\ref{fig3} illustrates that for $\lambda = 0.2$,
this asymptotic region is not yet reached for $Y \leq 5-7$, which is the range in which
the bulk of the currently known experimental data lie. 

To further illustrate the slow but steady convergence in shape between the MLLA results
for single-parton distributions and those of the coherent branching formalism, we have 
plotted in Fig.~\ref{fig5} the evolution of the peak position of these distributions as a function
of $Y$. We see that the peak positions in both approaches differ largely for $Y\leq 5-7$, 
consistent with our earlier remarks. In  general, the gluon jets split more than quark jets
and thus show a softer distribution and a peak at larger value of $l_{\rm max}$. This 
difference, however, decreases with increasing $Y$, as does the difference with the
peak position in the MLLA approach. The value of $l_{\rm max}/Y$ decreases slowly
with $Y$, but even for the largest values explored here (i.e. $Y=30$ corresponding to
the jet energy of $E \sim 10^{12}$~GeV), the position of the peak differs significantly 
from the DLA result $l_{\rm max}/Y = 1/2$.

\subsection{Total multiplicity and energy conservation}
\label{sec3b}
The single-parton distributions inside quark and gluon jets can be characterized
by their moments. The zeroth and first moments of the inclusive 
single-parton distributions are of particular interest, since they define the total parton 
multiplicity in a quark or gluon jet,
\begin{figure}[t]
\begin{center}
\includegraphics[height=14.5cm, angle=0]{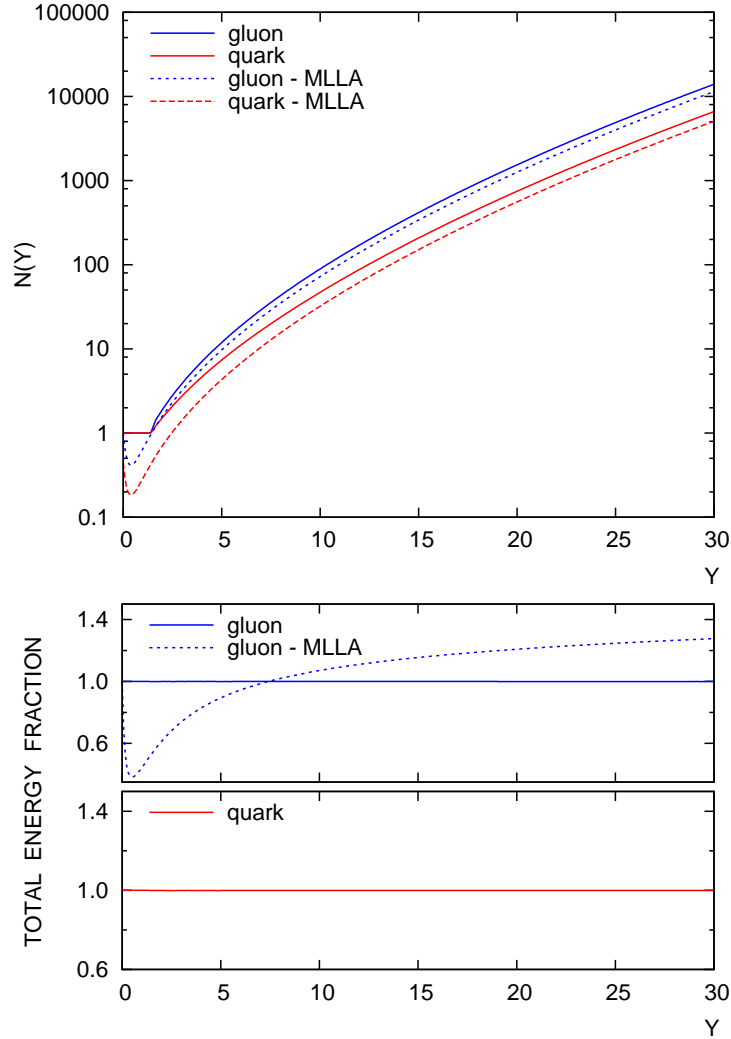}
\caption{Upper plot: $Y$-dependence of the total parton multiplicity (\protect\ref{3.2}) in a quark or 
gluon jet, calculated in the coherent parton branching formalism~(\protect\ref{eq:evoleqs})
(straight lines), and in the MLLA approach (\protect\ref{eq:mlla}) (dashed lines) for $\lambda = 0.2$. 
Lower plot: The ratio (\protect\ref{3.3}) of the total jet 
energy contained in the single parton distribution to the initial jet energy. 
The upper panel is for 
gluon jets, the lower one for quark jets, $\lambda = 0.2$. }
\label{fig6}
\end{center}
\end{figure}
%
%
\be
N_{q,g} (Y) = \int\! \! dl\, D_{q, g} (l, Y)\, ,
\label{3.2}
\ee
and the total energy fraction inside a quark or gluon jet 
\be
\frac{1}{E}\sum_{\rm partons} E_p(Y) = \int\! \! dx\, xD_{q, g} (x, Y)\, .
\label{3.3}
\ee
For an energy-conserving evolution of $D_{q, g}$, the first moment of the single-parton
distribution must yield the total jet energy $E$, and the fraction (\ref{3.3}) must equal
one. The bottom panel of Fig.~\ref{fig6} illustrates, that this is satisfied for the
coherent branching formalism. Since we know that the evolution equations (\ref{eq:evoleqs})
conserve energy, this is a cross check of the numerical accuracy of our solution.
We note that the value of the first moment (\ref{3.3}) is determined largely by 
$D_{q, g}(x, Y)$ in the region of large $x$. However, parametrically, the 
evolution equations (\ref{eq:evoleqs}) have been derived for the region of small $x$,  
where the logarithm $l$ is large. Fig.~\ref{fig6} also illustrates that the MLLA approach 
does not conserve energy exactly. So, on kinematic grounds, the coherent branching
formalism is certainly preferable.  We note, however, that for large values of $Y$, as 
seen in Fig.~\ref{fig4}, the region 
around the peak of the distribution $D_{q, g}$ will be at $x$ values, which are much softer
than the region of $x$ space which contributes dominantly to (\ref{3.3}). For this reason, 
a formalism which does not conserve energy may account accurately for the functions 
$D_{q, g}(x, Y)$ in a wide region around the peak, if $Y$ is sufficiently large.

Within the MLLA approach, one customarily relates by a prefactor the single-parton distributions 
in quark and gluon jets, so that $G(x,Y) = N_c/C_F\: Q(x,Y) $. This relation is 
valid to DLA accuracy
and leads to the first moment (\ref{3.3}) of a quark jet equal to the total energy 
faction of a gluon jet multiplied by $4/9$. We did not include this result in Fig.~\ref{fig6},
since it does not add further information.  

We now turn to the total parton multiplicities within a quark or a gluon jet, plotted 
in the upper panel of Fig.~\ref{fig6}. These multiplicities satisfy evolution equations for integrated
parton distributions, which are much simpler than the set of equations (\ref{eq:evoleqs}),
and which have been studied numerically~\cite{Lupia:1997in, Lupia:1997bs}. We checked that 
we reproduce these results, if we adopt the same modifications to the coherent branching 
formalism~(\ref{eq:evoleqs}). We also tested the numerical accuracy of our routines for 
$D_{q, g}(x, Y)$, by comparing the integral (\ref{3.2}) to the result of the simpler evolution
equations for multiplicities, and we found perfect agreement. 
Within MLLA, the multiplicities for the limiting case ($\lambda = 0$) and at large values of $Y$ are known to rise like  \cite{BasicsQCD,Mueller:1982cq,Dremin:1994bj}
\begin{equation}
	N_{\rm MLLA}(Y) \sim  Y^{-a_1/(2 \beta)+1/4}\, e^{2\, \sqrt{Y/\beta}}\,.
\end{equation}
This strong rise of the multiplicity is seen in Fig.~\ref{fig6}. At large $Y$, the rise is the
same in the
MLLA approach and in the coherent branching formalism. In comparison to the 
MLLA approach, the multiplicities obtained in the coherent branching formalism are a 
factor $\sim 1.24$ for gluon jets and a factor $\sim 1.30$ higher for quark jets at large $Y$.
This is consistent with our observation in Fig.~\ref{fig4} and indicates that at large $Y$, 
the distributions obtained in both approaches can be made to coincide by an 
approximately $Y$-independent multiplicative factor.

\subsection{Dependence of $Q(x,Y)$ and $G(x,Y)$ on $\lambda$}
\label{sec3c}

\begin{figure}[t]
\begin{center}
\includegraphics[height=13.0cm, angle=0]{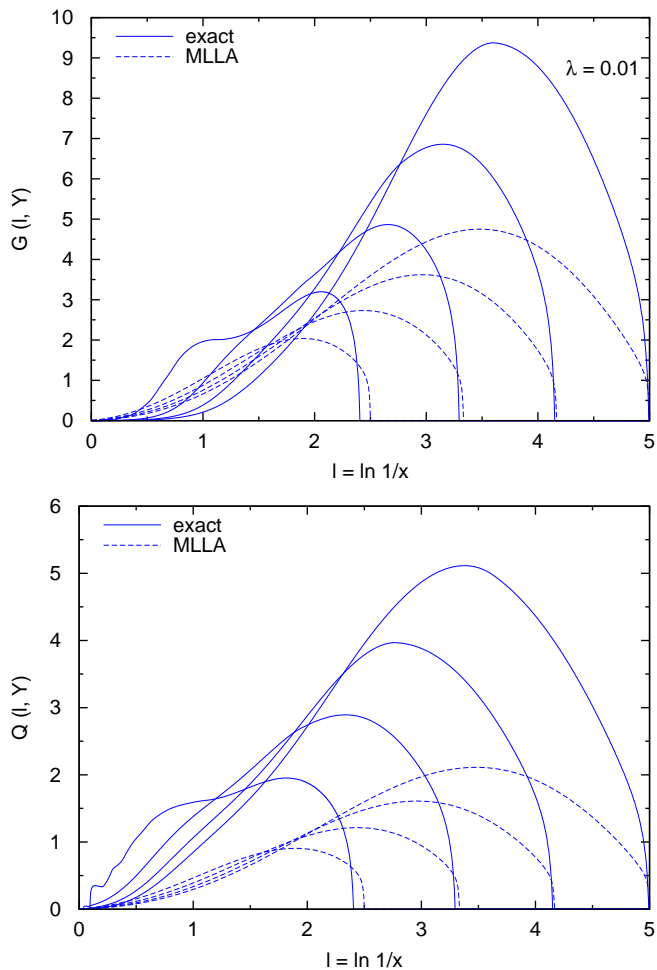}
\caption{Same as Fig.~\protect\ref{fig3} but for $\lambda = 0.01$.}
\label{fig7}
\end{center}
\end{figure}

The numerical results shown in the previous subsections ~\ref{sec3a} and 
~\ref{sec3b} were obtained for $\lambda = 0.2$. Here, we discuss the 
dependence of these results on $\lambda$. From the point of view of
data comparison, it is natural to regard the jet energy $E$
and the hadronization scale $Q_0$ as the two independent variables, 
while $\Lambda_{\rm QCD}$ is fixed. Specifying the $Y$-range of the 
evolution and the value of $\lambda$ amounts to specifying the variables $E$ and $Q_0$.
On the other hand, regarding the values for the jet energy and the $Y$-range 
of the evolution as given, decreasing $\lambda$ amounts to increasing 
$\Lambda_{\rm QCD}$. As a consequence, a smaller value of $\lambda$
is expected to lead to a more 'violent' evolution (i.e. more parton branchings) and 
it will thus lead to a higher jet multiplicity within the same $Y$-interval. 

\begin{figure}[t]
\begin{center}
\includegraphics[height=13.0cm, angle=0]{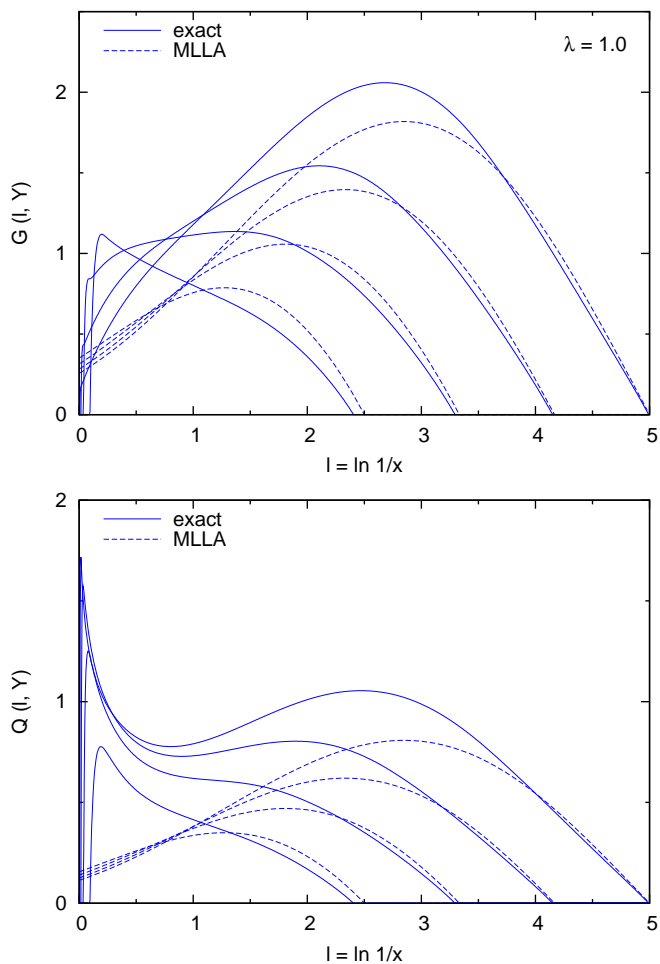}
\caption{Same as Fig.~\protect\ref{fig3} but for $\lambda = 1.0$.}
\label{fig8}
\end{center}
\end{figure}

This general expectation is confirmed by the curves in Figs. ~\ref{fig7} and \ref{fig8},
which show the distributions of all partons in a quark or a gluon jet for
evolution over small $Y$-intervals up to $Y\leq 5$.  
In particular, one sees that
after evolving for a few units in $Y$, the parton distributions for $\lambda = 1.0$
are still peaked at very large values of $x$, where $\ln 1/x \ll 1$. 
Qualitatively, the 
structures at large $x$ are due to the fact,
that for small $Y$ the phase space for evolution shown in Fig.~\ref{fig:cutoff} 
excludes evolution to very small values $x$, and the driving term of the
evolution is the Sudakov form factor at $x=1$. Upon evolution in $Y$, this 
Sudakov form factor decreases faster for the more violent evolution with $\lambda = 0.01$
than for less violent one with $\lambda = 1.0$, as can be seen from Fig.~\ref{fig2}. 
For this reason, the parton distributions in Fig.~\ref{fig7}, which were
calculated with $\lambda = 0.01$, decrease under evolution in $Y$ much faster 
in the region of larger $x$, than distributions evolved with larger values of $\lambda$, shown in Figs.~\ref{fig3} and \ref{fig8}. 
The same qualitative features (though quantitatively less pronounced)  are found in the 
MLLA approach, where for small $Y$ the continuous part of the single parton distribution
differs significantly from zero at large $x$, in particular for the case $\lambda = 1.0$ of a less 
violent evolution. 

For $\lambda = 0.01$ and $\lambda = 1.0$, we have calculated the single-parton 
distributions inside quark and gluon jets for evolution up to $Y=30$ (results not shown). 
In this asymptotic regime, similarly to the case $\lambda=0.2$, the results obtained from the coherent branching formalism 
and the MLLA approach can be made to coincide by a simple adjustment of 
the norm. For $\lambda = 1.0$, the MLLA results are at large values of $Y$ a factor  
1.11 smaller for gluon jets and a factor 1.16 smaller for quark jets. For $\lambda = 0.01$ 
the corresponding factors are 1.72 for gluon jets and 1.80 for quark jets.
As noted before in studies of total jet multiplicities \cite{Lupia:1997in, Lupia:1997bs}, the coherent branching
formalism does not allow for a limiting curve, $\lambda \to 0$, while the MLLA approach
does. Our numerical findings are consistent with this observation and indicate that
for decreasing $\lambda$, the difference in norm between both approaches increases
slowly but without bound.  

\begin{figure}[t]
\begin{center}
\includegraphics[height=8.1cm, angle=-90]{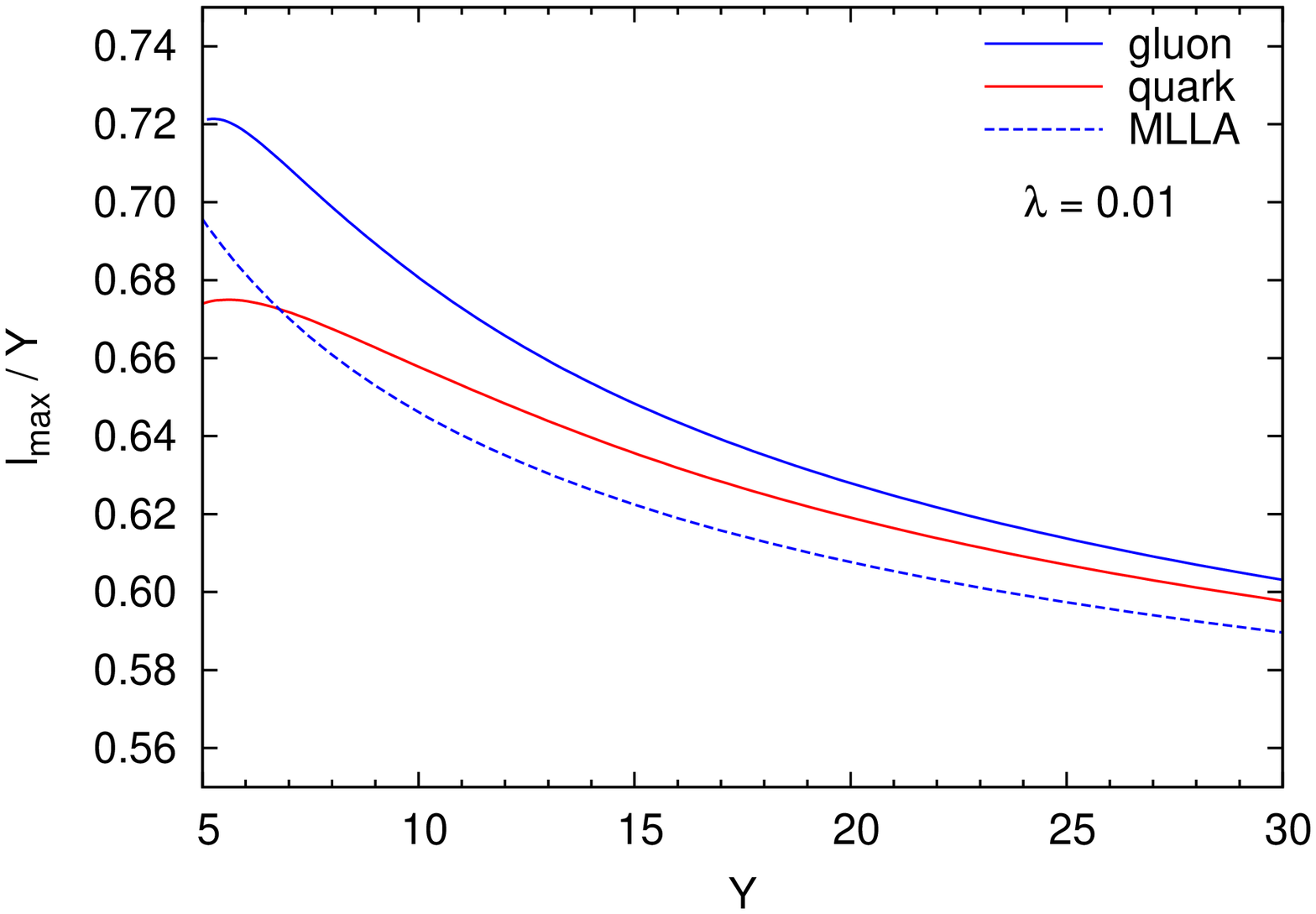}
\includegraphics[height=8.1cm, angle=-90]{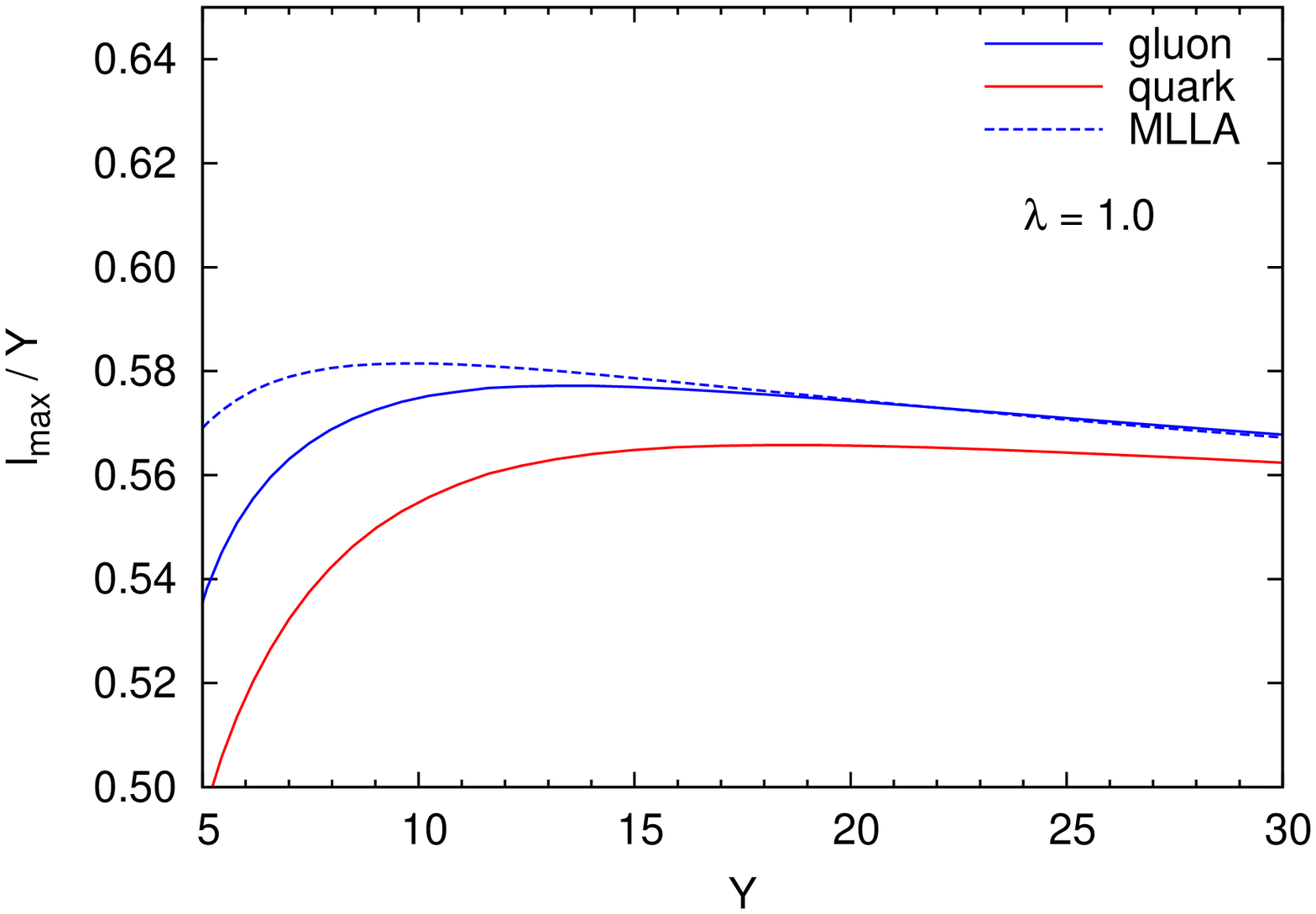}
\caption{Same as Fig.~\protect\ref{fig5} for $\lambda = 0.01$ and $\lambda = 1.0$.}
\label{fig9}
\end{center}
\end{figure}

For a more violent evolution, i.e. for smaller $\lambda$, 
single parton distributions increase in multiplicity and become 
softer. As a consequence, the peak position of the distributions is expected to shift to 
larger values of $\ln 1/x$ as $\lambda$ decreases. This expectation is confirmed in 
Fig.~\ref{fig9}, which shows the peak positions of single-parton distributions for 
$\lambda = 0.01$ and $\lambda = 1.0$.

\subsection{Matching MLLA to the coherent branching formalism}
\label{sec:maching}
As shown in Figs.~\ref{fig3}, ~\ref{fig7} and ~\ref{fig8}, single parton distributions
calculated in the MLLA approach differ for $Y \leq 5$ both in norm and in shape from 
those calculated in the coherent branching formalism. Besides the normalization,
 the spectra depend on two additional parameters, namely $Q_0$ and $\Lambda_{\rm QCD}$ or equivalently $Y$ and $\lambda$. Here we ask to what extent a 
variation of the three parameters is sufficient to make results from the 
coherent branching formalism coincide with a given result of the limiting fragmentation 
function ($\lambda \to 0$) obtained in the MLLA approach. One motivation for this 
study comes from the observation that the MLLA limiting fragmentation function is at the 
basis of many phenomenologically successful comparisons. So, even if the 
comparison of QCD predictions for single-hadron distributions in jets is not restricted
to the calculation of single-parton distributions and involves difficult issues in the modeling
of hadronization, one wonders to what extent the shape of partonic distributions
in the two calculational schemes can be made to agree by a suitable choice of parameters. 

\begin{figure}[t]
\begin{center}
\includegraphics[height=17.0cm, angle=-90]{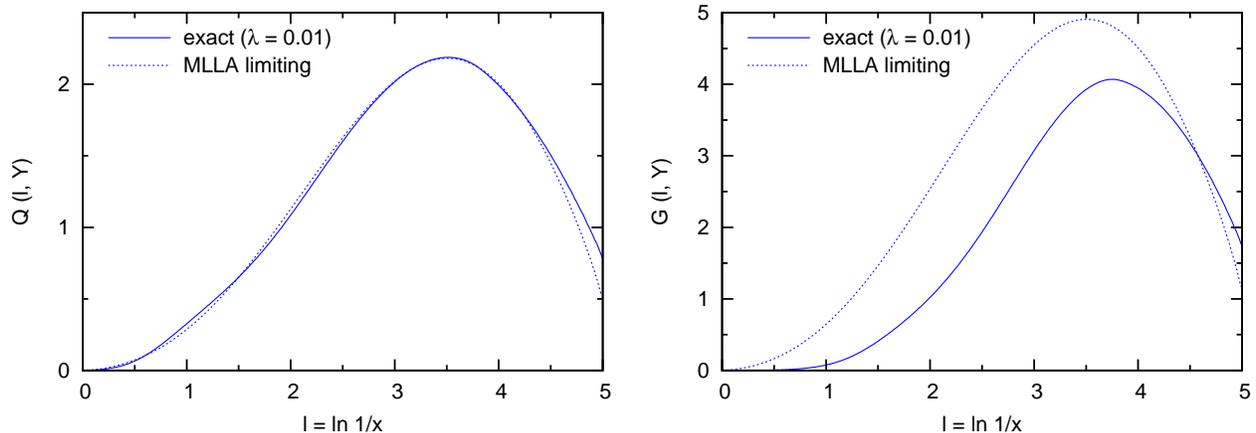}
\caption{The continuous parts of single parton distributions. Dashed lines:
the MLLA limiting fragmentation for $Y=5.0$. Solid lines:
results for the coherent branching formalism with a choice of $\lambda = 0.01$, $Y=5.2$  and
norm adjusted to $0.405$ such that the distributions in quark jets almost match.}
\label{fig10}
\end{center}
\end{figure}
\begin{figure}[t]
\begin{center}
\includegraphics[height=9.0cm, angle=-90]{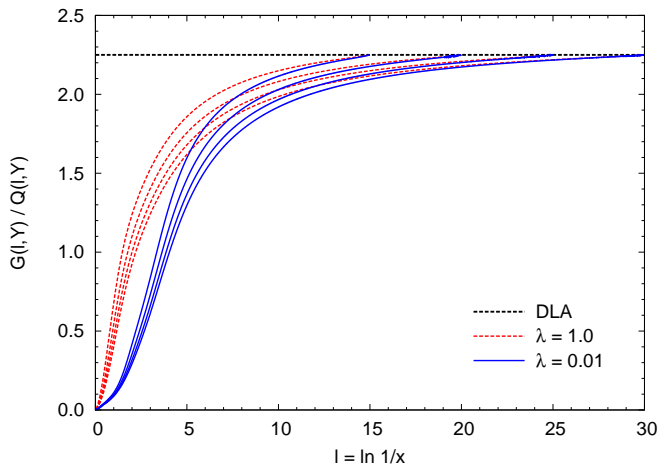}
\caption{The ratio $G(l,Y)/Q(l,Y)$ in the coherent branching formalism for two different 
values of $\lambda$ and evolution up to different $Y=15$, $20$, $25$ and $30$, from
top to bottom. 
By construction, MLLA results assume the DLA relation (\protect\ref{2.**}). }
\label{fig11}
\end{center}
\end{figure}

Fig.~\ref{fig10} shows an example of the extent to which this is possible. Adjusting
the norm, the evolution time, $Y$, and the value of $\lambda$, the single-parton distributions in a quark jet can 
be made to coincide almost for both formalisms at $Y=5$. More generally, one sees
from Figs.~\ref{fig5} and ~\ref{fig9} that the peak position of the single-parton 
distributions in either a quark or a gluon jet can always be adjusted to agree between 
both approaches by varying $Y$ and $\lambda$ in the coherent branching formalism. The total 
yield can then be adjusted by the norm. 

On the other hand, it is not possible for $Y\sim5$ and with the same choice of 
$\lambda$ to get results of both formalisms to coincide for the distributions in both 
quark and gluon jets . This is illustrated by the example given in Fig.~\ref{fig10}, where 
marked differences persist between the distributions in gluon jets, once the distributions 
in quark jets have been adjusted. This conclusion is supported more generally by
Fig.~\ref{fig11}, which shows the ratio of single-parton distributions of a gluon and a 
quark jet. Within the MLLA approach, these distributions vary in norm but not in shape,
since the ratio $G(l,Y)/Q(l,Y)$ is fixed to the DLA value $N_c/C_F$. On the contrary, within the coherent
branching formalism, the distributions in a quark jet and a gluon jet vary both in
norm and shape. In this latter case, the deviation from $N_c/C_F$ is more
pronounced in the region of relatively large $x$, and upon evolution in $Y$ this region 
becomes disentangled from the region around the peak of the distribution. In short, 
while there are marked differences between both formalisms at all $Y$, for 
sufficiently large $Y$ these differences die out in the region of large $l\sim {\cal O}(Y)$ to which the
coherent branching formalism and the MLLA approach are tailored. 

\section{Identified parton distributions in quark and gluon jet}
\label{sec:idpartons}

So far, we have discussed the distribution of {\it all} partons within a quark or a gluon jet. 
In the coherent branching formalism, we can study separately the distribution of \emph{quarks}
and \emph{gluons} within a quark or a gluon jet. We characterize these distributions by subscripts.
For instance, $Q^g(l,Y)$ is the single-gluon distribution in a quark jet. As explained in
subsection~\ref{sec:init}, single-quark distributions are obtained by evolving from the
initial conditions  (\ref{eq:init1}) and single-gluon distributions are obtained by evolving
from the initial condition (\ref{eq:init2}). The phase space constraints imply that there is no
evolution in the coherent branching formalism up to $Y=\ln 4$. 
Since the continuous parts of the distributions vanish by definition at $Y=\ln 4$, the initial conditions \eqref{eq:init1} and \eqref{eq:init2} translate directly to the initial conditions for the Sudakov form factors
\be
\begin{array}{lll}
S_g(Y=\ln 4)  =  0\, ,  \qquad &
S_q(Y=\ln 4)  =  1\, ,  \qquad &
\qquad {\rm if\ only\ quarks\ are\ counted,}  \\
S_g(Y=\ln 4)  =  1\, , \qquad &
S_q(Y=\ln 4)  =  0\, , \qquad &
\qquad {\rm if\ only\ gluons\ are\ counted.}
\end{array}
\ee
It is straightforward to check that if a Sudakov factor equals zero at $\ln 4$, then it stays zero at 
any value of $Y$.

\begin{figure}[t]
\begin{center}
\includegraphics[width=8.0cm]{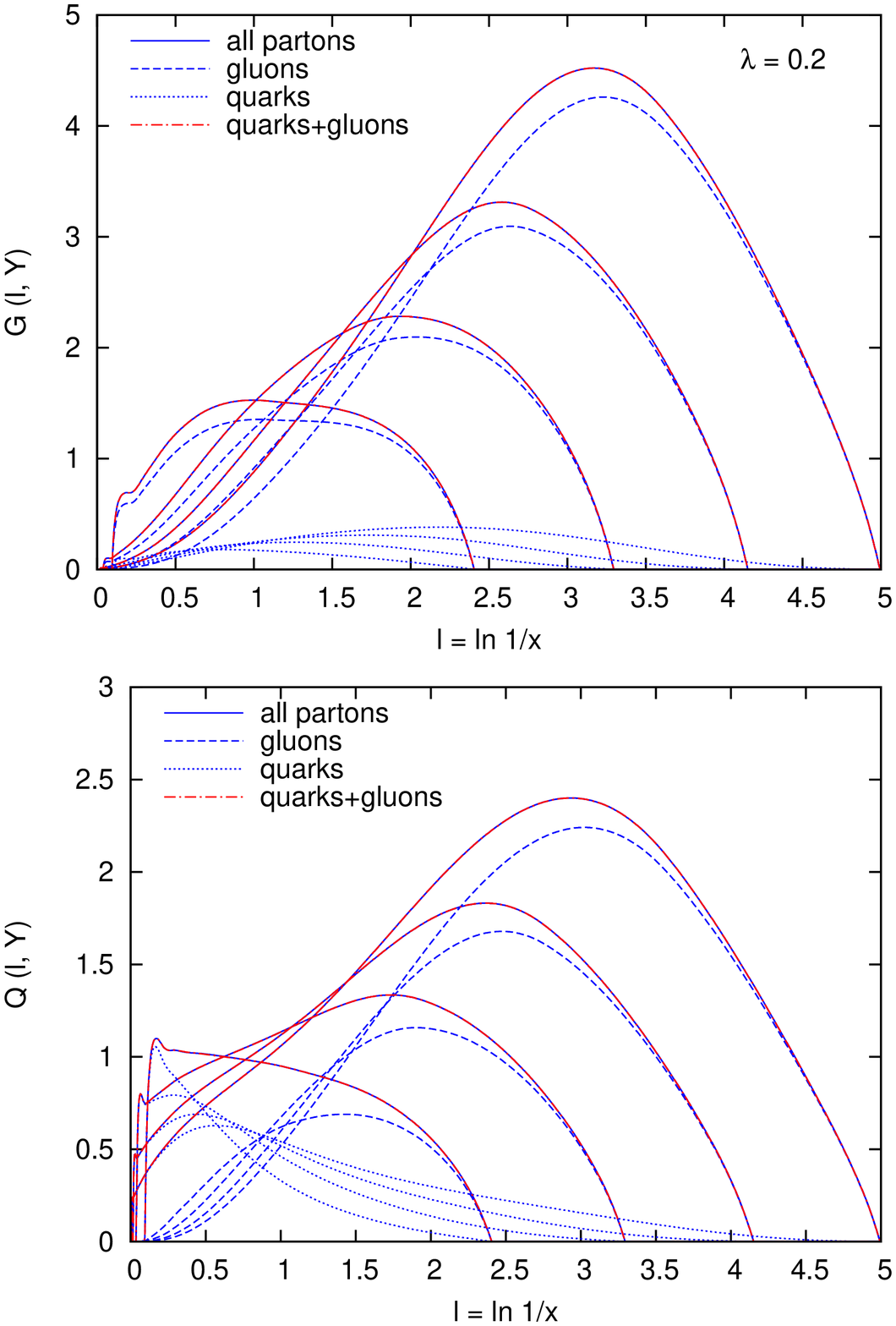}
\includegraphics[width=8.0cm]{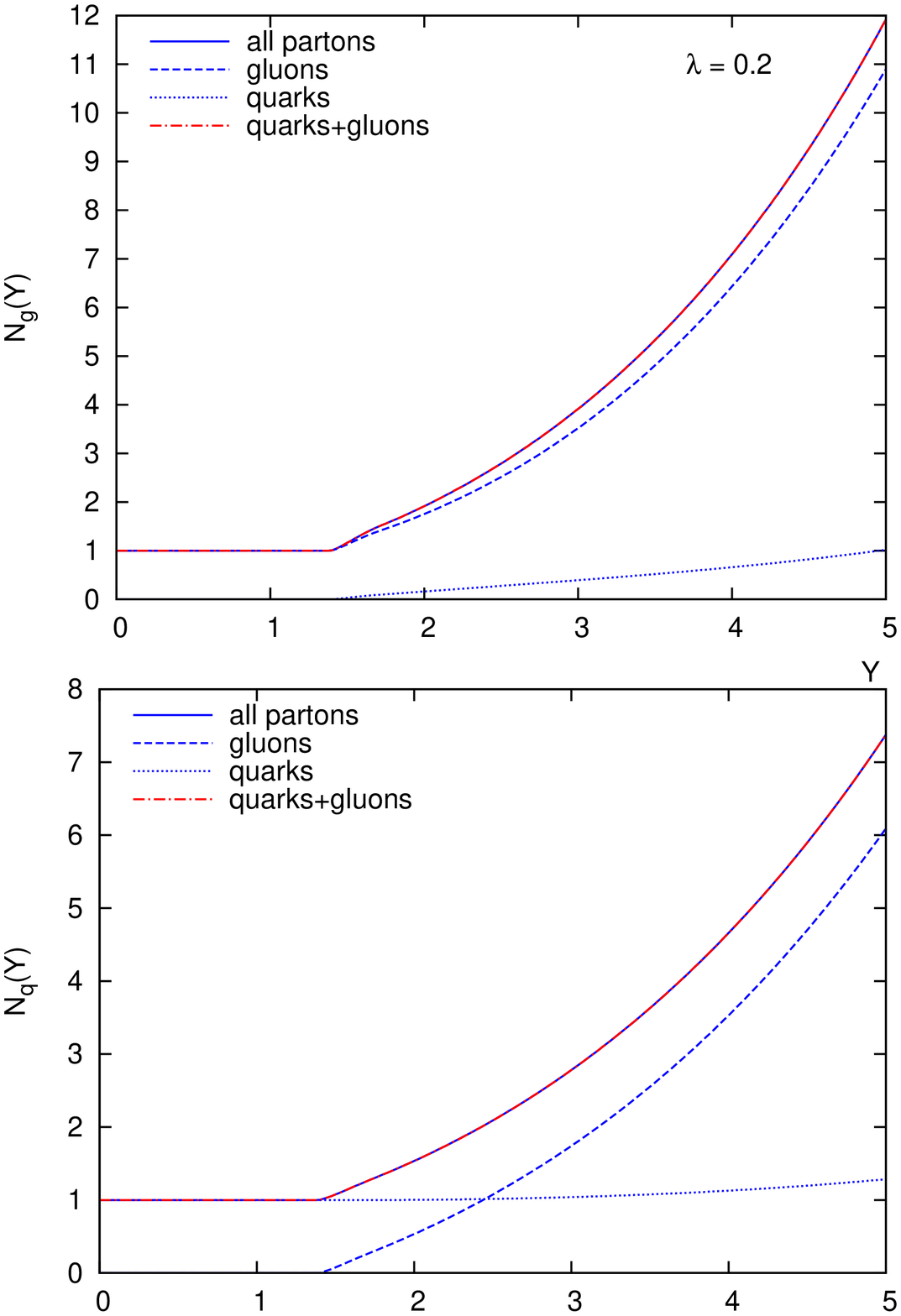}
\caption{Left hand side: The continuous part of the single-parton distributions inside a gluon jet
(upper panel) and a quark jet (lower panel). Different curves denote the single-quark
(dashed) and single-gluon (dotted) distributions inside a jet, the explicit sum 
of these two distributions (dash-dotted), and its comparison to the distribution of
all partons in a jet (solid) obtained by evolving the initial condition (\protect\ref{eq:init3}). 
The two latter curves coincide, as expected. Right hand side: the 
$Y$-dependence of the total multiplicity of quarks, gluons and all partons calculated 
for the cases shown on the left hand side.}
\label{fig:idpartons}
\end{center}
\end{figure}

In general, the distribution of quarks inside a jet is harder than that of gluons.
This is a consequence of the fact that gluons are more likely to split.
In the left panel of Fig.~\ref{fig:idpartons}, which shows the identified parton 
spectra in jets for $\lambda=0.2$,
this is clearly seen for all values of $Y$. A particularly pronounced feature is seen in the
single-quark distribution for quark jets: up to $Y=5$, this distribution peaks at very
large $x$ values corresponding to $l < 0.5$. The integrated total quark multiplicity
in this distribution, shown in the right panel of Fig.~\ref{fig:idpartons},  grows slowly but remains of order unity during this evolution, 
indicating that this single quark distribution follows closely the momentum distribution
of the evolved parent quark. This parent quark evolves via $q \to q\, g$ splittings
predominantly by emitting soft gluons. For this reason, the quark distribution 
remains hard and the total multiplicity in the quark jet is largely dominated by
subsequent $g \to g\, g$ splittings. It is in precisely this sense that the enhancement
at large-$x$ is a remnant of the $\delta$-function in the initial condition, which gradually
becomes negligible in the evolution to larger $Y$. 

We have emphasized repeatedly that the approximations involved in deriving the
coherent branching formalism do not guarantee an accurate description of the
large-$x$ region. We note, however, that the reasons which we have identified for the
pronounced large-$x$ enhancement of $Q^q(x,Y)$ in Fig.~\ref{fig:idpartons} do not refer
to particular features of an ${\cal O}(\sqrt{\alpha_s})$ approximation, but emerge 
solely from generic features of the branching of quarks and gluons. As a consequence,
we expect that these features, though observed outside the strict region of validity of
the formalism explored here, correspond to physical reality and persist in a more complete 
formulation of the problem, which is accurate in the large-$x$ region. Indeed, in Monte Carlo 
event generators, such an enhancement is seen on the level of partonic 
distributions \cite{Sjostrand:2006za,Marchesini:1991ch,Gleisberg:2003xi}. 
In these event generators, it is generally the hadronization 
mechanism which connects the color of the leading quark to the rest of the event,
and which thus leads to a hadronic single-inclusive distribution which is much 
softer than the partonic one. 

For the distribution of quarks within a gluon jet, the situation is clearly different since
the splitting function for $g \to q\, \bar{q}$ does not give rise to a logarithmic enhancement.
For this reason, the total quark multiplicity inside a quark or a gluon jet rises only slowly.
For gluon jets as well as for quark jets, the jet multiplicity is ultimately dominated by gluons. The
distribution of the hardest gluon in the first $g \to g\, g$ splitting process is likely to 
contribute significantly to the large values which $G(l,Y)$ shows in the large $x$ 
region (say $l<1.0$) for small values of $Y$. Upon further evolution in $Y$,
this structure disappears faster in $G(l,Y)$ than in $Q(l,Y)$, since it is dominated in the
first case by gluons, which split more readily than quarks. 

\section{Conclusion}
\label{sec:conclusion}

We have calculated single parton distributions inside quark and gluon jets by solving the
full evolution equations in the coherent branching formalism. For
sufficiently large jet energies, corresponding to $Y > 5 - 10$, results obtained from the MLLA
and from the coherent branching formalism agree very well in shape. They differ
in this asymptotic regime solely by a factor of order unity in norm, and this factor
depends weakly on the choice of hadronization scale. For smaller values of the jet
energy ($Y \leq 5-7$), however, both formalisms show marked differences in norm and
shape. We have characterized these differences in detail in sections~\ref{sec:numres} 
and ~\ref{sec:idpartons}.

We recall that, parametrically, both MLLA and the coherent branching formalism are valid
for sufficiently large $Y$ in a kinematic region in which $l \sim {\cal O}(Y)$. As a consequence,
the differences observed between both approaches for $Y \leq 5-7$ can be regarded as
preasymptotic effects, which persist solely outside the strict region of validity of both approaches.
If one adopts this view, then our study has contributed to delineating the region of
quantitatively reliable applicability of both approaches, namely $Y > 5-7$. 
In this context, we note that for $Y<5-7$, single-parton distributions in both approaches
cannot be made to coincide by a suitable choice of parameters for both quark and gluon jets,
while they can be made to coincide for one of them, see section~\ref{sec:maching}.

On the other hand, the MLLA limiting spectrum has been compared successfully to a 
large data set of single inclusive hadronic distributions for jet energies, which lie in the
range $Y \leq 5-7$. These data comparisons required the assumption of local parton hadron
duality and, by their success, gave support to this assumption. In the present work, we have
found that for $Y \leq 5 - 7$, distributions calculated in the coherent branching formalism 
show a significant enhancement at large $x$, which is much less pronounced and vanishes
much faster with $Y$ in the MLLA approach. Also, this enhancement has not been
observed in the measured hadronic distributions. 

Although these findings were made in the region of small $Y$ and large $x$, which lies 
clearly outside the region of quantitative applicability of the formalisms in question, we have 
argued in section~\ref{sec:idpartons} that the structures found at large $x$ in the coherent
branching formalism emerge solely from generic properties of the branching of
quarks and gluons and should thus persist in a more complete formulation of the problem,
which is reliable at small $Y$ and large $x$. We note that the marked difference between the 
shape of partonic distributions calculated in the coherent branching formalism 
and the shape of the measured hadronic distributions does not necessarily invalidate the 
coherent branching formalism in the range of $Y\sim 3-7$. Rather, this
difference may be indicative of a non-trivial dynamics of hadronization, which is 
characteristically different from a simple one-to-one mapping of partons into hadrons
at the scale $Q_0 \sim {\cal O}(\Lambda_{\rm QCD})$. Indeed, since the hardest partonic
components in a jet are connected to the softer ones by color flow, and since
hadronization is ultimately sensitive to the color structure of the event, the dynamics
of hadronization may lead to a significant softening of the single inclusive partonic
large-$x$ distributions. Such features are realized at least in some models of hadronization,
such as the Lund string model. If one adopts this view, then our study may be regarded 
as a contribution to the question to what extent the measured inclusive hadronic
distributions inside jets lend support to a specific picture of hadronization.

\medskip
\section*{Acknowledgments}
\medskip
We thank Yuri Dokshitzer, Peter Richardson, Peter Skands, and Bryan Webber for 
helpful discussions at various stages during this work. We are particularly indebted
to Krzysztof Golec-Biernat, who was a valuable discussion partner on all aspects
of this study. S.S. is grateful to the CERN Theory Group for support and warm hospitality,
and he acknowledges support from the Foundation for Polish Science (FNP) and a 
grant of the Polish Ministry of Science, N202 048 31/2647 (2006-08).


\end{document}